\documentclass[usegraphicx,usenatbib]{mn2e}
\usepackage{times}
\usepackage{rotating}
\def\mpc{h^{-1} {\rm{Mpc}}}
\def\kms {\rm{km~s^{-1}}}
\def\apj {ApJ}

\def\apjs {ApJS}
\def\aj {AJ}
\def\aap {A\&A}
\def\mnras {MNRAS}

\begin{document}
\title [Group dynamics and luminosity content]
{The dynamical state of galaxy groups and their luminosity content}
\author[Mart\'inez \& Zandivarez]
{H\'ector J. Mart\'inez$^{1,2}$\thanks{julian@oac.uncor.edu} \& Ariel Zandivarez$^{1,2,3}$\\
$1$ Instituto de Astronom\'{\i}a Te\'orica y Experimental, IATE, CONICET, Laprida 854, X5000BGR, 
C\'ordoba, Argentina\\
$2$ Observatorio Astron\'omico, Universidad Nacional de 
C\'ordoba, Laprida 854, X5000BGR, C\'ordoba, Argentina\\
$3$ Instituto de Astronom\'ia, Geof\'isica e Ciencias Atmosfericas, IAG, USP, Rua do Mat\~ao 1226,
S\~ao Paulo, Brazil}
\date{\today}
\pagerange{\pageref{firstpage}--\pageref{lastpage}} 
\maketitle
\label{firstpage}
\begin{abstract}
We analyse the dependence of the luminosity function of galaxies in groups (LF)
on group dynamical state. We use the Gaussianity of the velocity distribution
of galaxy members as a measurement of the dynamical equilibrium of groups identified
in the SDSS Data Release 7 by Zandivarez \& Mart\'inez 2011. 
We apply the Anderson-Darling goodness-of-fit test to distinguish between groups
according to whether they have Gaussian or Non-Gaussian velocity distributions, i.e.,
whether they are relaxed or not.
For these two subsamples, we compute the $^{0.1}r-$band LF as a function of group virial mass 
and group total luminosity. For massive groups, ${\mathcal M}>5 \times 10^{13} \ M_{\odot} \ h^{-1}$, 
we find statistically significant differences between the LF of the two subsamples:
the LF of groups that have Gaussian velocity distributions have a brighter characteristic 
absolute magnitude ($\sim0.3$ mag) and a steeper faint end slope ($\sim0.25$).
We detect a similar effect when comparing the LF of bright ($M^{group}_{^{0.1}r}-5\log(h)<-23.5$) 
Gaussian and Non-Gaussian groups.
Our results indicate that, for massive/luminous groups, the dynamical state of the
system is directly related with the luminosity of its galaxy members.
  
\end{abstract}
\section{Introduction} 
Group environment plays a key role in the evolution of the overall 
galaxy population in the Universe. Many works in the literature have studied galaxies
in groups to understand how this particular environment affects galaxies
and their properties (e.g. \citealt{m02,eke04b,balogh04,zmm06,weinmann06,yang09,robot10}).
The understanding of how the different processes act upon galaxies requires not only
a characterisation of how galaxy properties are related to the environment but also 
to their motion therein. 
The action of the different physical mechanisms depend on the dynamics of galaxies within systems
(e.g. \citealt{yepes91,fusco98,abadi}).
There is evidence that some properties of galaxy systems 
are closely related to galaxy dynamics (e.g. \citealt{whit93,adami98,biviano02,lares04,ribeiro10}).

A possible way to characterise the dynamical state of a galaxy group is analysing its velocity
distribution. It is known that a Gaussian velocity distribution is indicative of 
a group in dynamical equilibrium, while departures from Gaussianity may indicate that perturbative
processes are working \citep{menci96,ad}.
However, there is a difficulty in determining whether a given velocity distribution differs 
significantly from Gaussian, mainly when studying smaller systems as galaxy groups with only a few
galaxy members.
In a recent work, \citet{ad} have demonstrated that a reliable distinction can be made between
Gaussian and non-Gaussian groups even for those with low group membership. They conclude
that the Anderson-Darling (A-D) goodness-of-fit test is the most reliable statistics to 
distinguish between relaxed and 
dynamically disturbed systems even for those with at least 5 galaxy members.
Therefore, this statistical method is a very suitable tool to analyse the internal dynamics of a system. 

Recently, \citet{zm11} (hereafter ZM11) used the Seventh Data Release of the
Sloan Digital Sky Survey (hereafter SDSS DR7; \citealt{dr7}) to identify groups of galaxies and
study several dependencies of the LF. They found that the characteristic magnitude brightens and 
the faint end slope becomes steeper as a function of mass. This change in the luminosity function 
is mainly due to the red spheroids and the varying number contributions of the different galaxy types. 
They also found evidence of luminosity segregation for massive groups.
Moreover, the mass trend of the LF is much more pronounced for groups located in low density regions. 
However, the effects of the internal dynamics of groups on the galaxy LF is an issue that has not 
been fully addressed. Therefore, in this paper we extend the work by ZM11 by 
studying the link between the LF and the dynamical state of groups by means of their galaxy member 
velocity distributions using the A-D test.
The layout of this paper is as follows. In section 2 we describe the
group sample. The analysis of the LFs is in section 3. 
Finally, in section 4 we discuss the results. 

\section{The sample}
For this work, we use the sample of groups constructed by ZM11. 
This sample has been identified in the Main Galaxy Sample (MGS; \citealt{mgs}) of SDSS DR7 
which comprises galaxies down to an apparent magnitude limit of $17.77$ in the $r$ band.
In ZM11, the group identification was performed following \citet{mz05}:
firstly, a standard Friends-of-Friends ({\em fof}) algorithm links MGS galaxies into groups; and 
secondly, an improvement of the rich group identification is performed by means of a second identification 
on galaxy groups which have at least ten members using a higher density contrast. The latter
is done in order to split merged systems or to eliminate spurious member detection (see \citealt{diaz05}). 
The method for estimating group centre positions was refined for groups with at least ten members 
using an iterative procedure developed by \citet{diaz05}. 
Due to the well known incompleteness of the MGS for ${\it r}<14.5$,
ZM11 excluded galaxies brighter than $14.5$. The linking parameters for the {\em fof} 
algorithm were set to have a transverse linking length 
which corresponds to a contour over-density of $\delta \rho/\rho=200$ and a 
line-of-sight linking length of $200~\kms$.
As in \citet{merchan02}, group virial masses were computed as
${\cal M}=\sigma^2R_{\rm vir}/G$,
where $R_{\rm vir}$ is the virial radius of the system, and $\sigma$ is the velocity dispersion of
member galaxies \citep{limber60}.  
The velocity dispersion $\sigma$ was estimated using the line-of-sight velocity dispersion $\sigma_{v}$,
$\sigma=\sqrt{3}\sigma_v$. To compute $\sigma_v$ we used the methods described by \citet{beers90},
applying the biweight estimator for groups with richness $N_{gal}\ge 15$ and the gapper
estimator for poorer systems.
The final group sample comprises 15,961 groups with at least 4 members,
adding up to 103,342 galaxies. The group sample has a median redshift, velocity dispersion, virial 
mass and virial radius of 0.09, $193 \ \kms$, $2.1\times10^{13} \  M_{\odot} \ h^{-1}$, 
and $0.9 \ \mpc$, respectively. 

\begin{figure}
\begin{center}
\includegraphics[width=70mm]{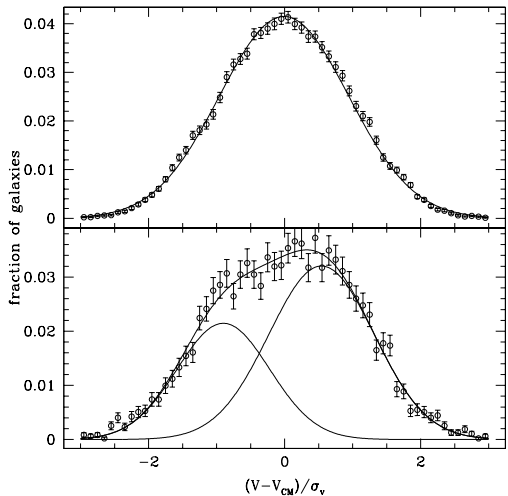}
\caption{
The normalised line-of-sight velocity distribution of galaxies in Gaussian 
({\em upper panel}) and Non-Gaussian ({\em lower panel}) groups.
Solid lines are the fitting functions describing the distributions (see text for details).
}
\label{fig1}
\end{center}
\end{figure}
\begin{figure}
\begin{center}
\includegraphics[width=70mm]{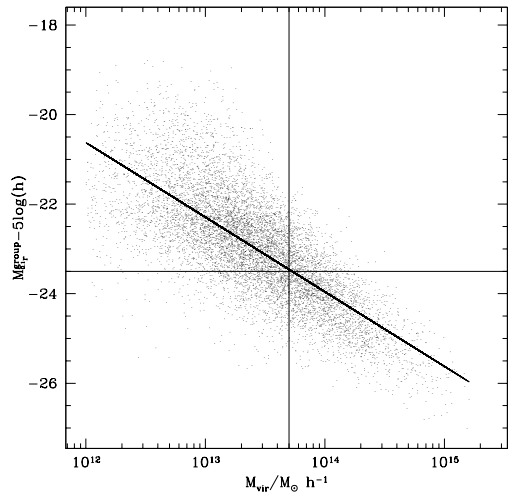}
\caption{
The group total absolute magnitude vs. virial mass for ZM11 groups which have at least 5 members. 
The {\em thick solid line} is the least square linear fit between $M^{group}_{^{0.1}r}$
and $\log({\mathcal M}_{vir})$.
{\em Vertical line},  ${\mathcal M}_{vir}=5\times 10^{13}M_{\odot}h^{-1}$, is the high mass cut-off,
while {\em horizontal line}, $M^{group}_{^{0.1}r}-5\log(h)=-23.5$, is the corresponding 
high luminosity cut-off obtained from the estimated linear relation and the high mass cut-off value.
}
\label{fig2}
\end{center}
\end{figure}

Galaxy magnitudes used throughout this paper are Petrosian, are in the AB system and 
have been corrected for Galactic extinction using the maps by \citet{sch98}. 
Absolute magnitudes have been 
computed assuming a flat cosmological model with parameters $\Omega_0=0.3$, 
$\Omega_{\Lambda}=0.7$ and $H_0=100~h~{\rm km~s^{-1}~Mpc^{-1}}$ 
and $K-$corrected using the method of \citet{blantonk}~({\small KCORRECT} version 4.1). 
We have also included evolution corrections to this magnitude following \citet{blantonlf}.
We have adopted a band shift to a redshift $0.1$ for the $r$ band (hereafter $^{0.1}r$), i.e. to 
approximately the mean redshift of the main galaxy sample of SDSS. 

\begin{figure}
\includegraphics[width=90mm]{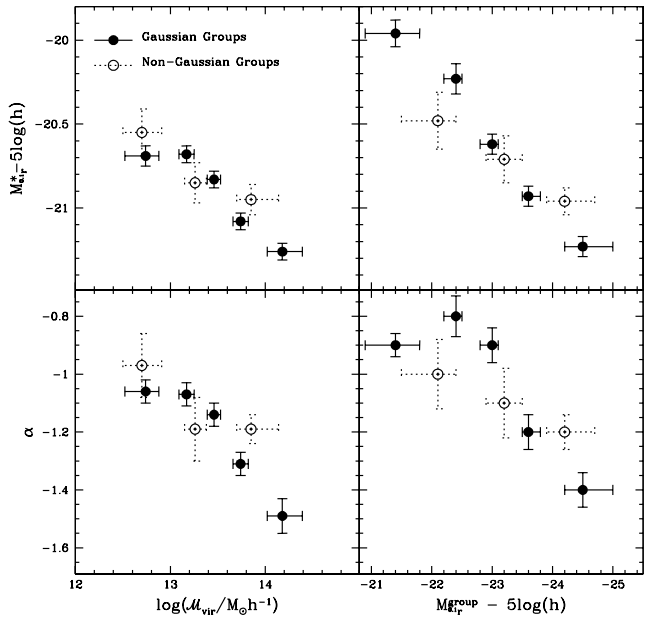}
\caption{STY best fitting Schechter function parameters in the $^{0.1}r$ band as 
a function of group mass ({\em left panels}) and group total absolute magnitude ({\em right panels})
for the subsamples of G ({\em filled circles}) and NG ({\em open circles})
groups. {\em Upper panels} show the characteristic absolute magnitude,
while {\em lower panel} show the variation of the faint end slope.
Vertical error bars are the projection of  1$\sigma$ joint error ellipse onto 
the $\alpha$ and $M^{\ast}$ axes. 
Horizontal error bars are the 25th and 75th percentiles in each mass/luminosity bin. 
}
\label{fig3}
\end{figure}

To distinguish between relaxed groups with Gaussian velocity distributions and groups 
with non-Gaussian dynamics we adopted the A-D goodness-of-fit test. 
In a recent study, \cite{ad} have demonstrated that the A-D is one of the most 
reliable and powerful tests to measure departures from an underlying Gaussian distribution.
This test does not require binning or graphical analysis of the data. 
From the outcome of the A-D test, we classify groups into two subsamples: the non-Gaussian
(NG) groups are those which have a confidence level above $90\%$  of not having a Gaussian
velocity distribution, while Gaussian (G) groups are those whose confidence level 
of not having a Gaussian velocity distribution is below $50\%$.
Since the A-D test is reliable when data sets have at least 5 points \citep{ad}, for the purposes
of this work we restrict the ZM11 sample to groups with at least 5 members. From the total of 9,387
groups with at least 5 galaxy members, 479 are classified as NG while 5,250 as G groups.
In Fig. \ref{fig1} we show, for our samples of G and NG groups, the stacked distribution of
the radial velocity of the galaxies ($V$) relative to their parent group radial velocity ($V_{CM}$) 
normalised to the group velocity dispersion ($\sigma_v$). 
We perform a fitting procedure to clearly describe the velocity distribution behaviours.
It can be seen ({\em solid lines}, Fig. \ref{fig1}) that the stacked velocity distribution of G groups
is well represented by a Gaussian function ({\em upper panel}), while the NG groups show
clear departures from a single Gaussian function, being well fit by the sum of two Gaussian
functions ({\em lower panel}). The small asymmetry in the velocity profile of NG groups is due to
the only 4 groups which have more than 90 members. Among them, 3 have left-skewed radial velocity distribution,
and 1 group has a right-skewed one. By excluding these large groups, the stacked radial velocity 
distribution of NG groups becomes very close to symmetrical, still non-Gaussian and well 
fit by the sum of two Gaussian functions displaced from each other.

\section{The lf of galaxies in groups}
This study is based on the analysis of the luminosity function of galaxies in groups as a function of 
a given group physical property, using the group subsamples defined in the 
previous section. As in ZM11, the system physical property adopted is the group virial mass.
For groups that are not in dynamical equilibrium, i.e. those classified as NG, 
the virial mass might not be a suitable measure of the system mass. Thus, comparing
the LF of G and NG groups as a function of mass can be thought as inappropriate.  
Thus, to complement the analysis of the LF we use also another group property which is known to be 
correlated with mass, the group total luminosity (e.g. \citealt{girardi00,pop05,dm05}). 
We compute group total absolute magnitudes following \citet{moore93} but using the mass
dependence of the LF of galaxies in groups as computed by ZM11.
We show in Fig. \ref{fig2} the group total absolute magnitudes of the ZM11 groups as a function
of their virial masses. It can be seen that there is a real correlation among these
parameters.

Similarly to the obtained by ZM11, in all cases, we find that
the Schechter parametrisation of the LF \citep{schechter76} is 
appropriate for describing the binned LF\footnote{The binned LF were computed using the 
$C^-$ method \citep{lb71,cho87}.}. 
Therefore, our findings below are expressed in terms of the values of the Schechter function 
shape parameters, $\alpha$ and $M^{\ast}$, only, which we compute using the STY method \citep{sty79}.

In Fig. \ref{fig3} we show the best fitting $\alpha$ and $M^{\ast}$ parameters of the 
$^{0.1}r-$band LFs of galaxies in G and NG groups as a function of their 
virial masses ({\em left panels}) and total absolute magnitude ({\em right panels}). 
We use as many bins as possible to probe the mass/absolute magnitude range while having
enough data points in each bin to produce reliable estimations of the LF.
We use 5 and 3 bins which include the same number of groups for the G and the NG
group samples respectively.

\begin{figure}
\includegraphics[width=90mm]{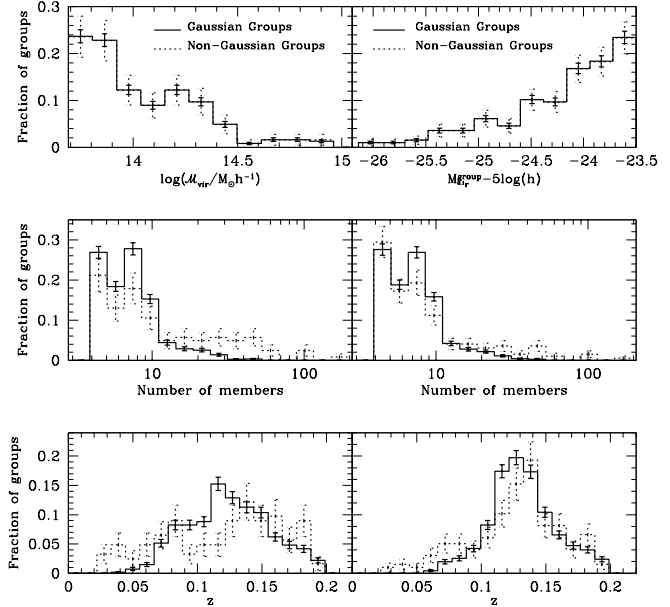}
\caption{
Groups in the single high mass/luminosity bins. {\em Left panels} correspond
to groups with ${\mathcal M}_{vir}>5\times 10^{13}M_{\odot}h^{-1}$ and {\em right panels}
to groups  brighter than $M^{group}_{^{0.1}r}-5\log(h)=-23.5$.
From {\em top} to {\em bottom} we show the normalised distributions of:
virial mass/total luminosity, number of members and redshift,
of G ({\em solid line}) and NG ({\em dotted line}) groups.
Error bars in each histogram are Poisson errors.
}
\label{fig4}
\end{figure}

\begin{figure*}
\includegraphics[width=170mm]{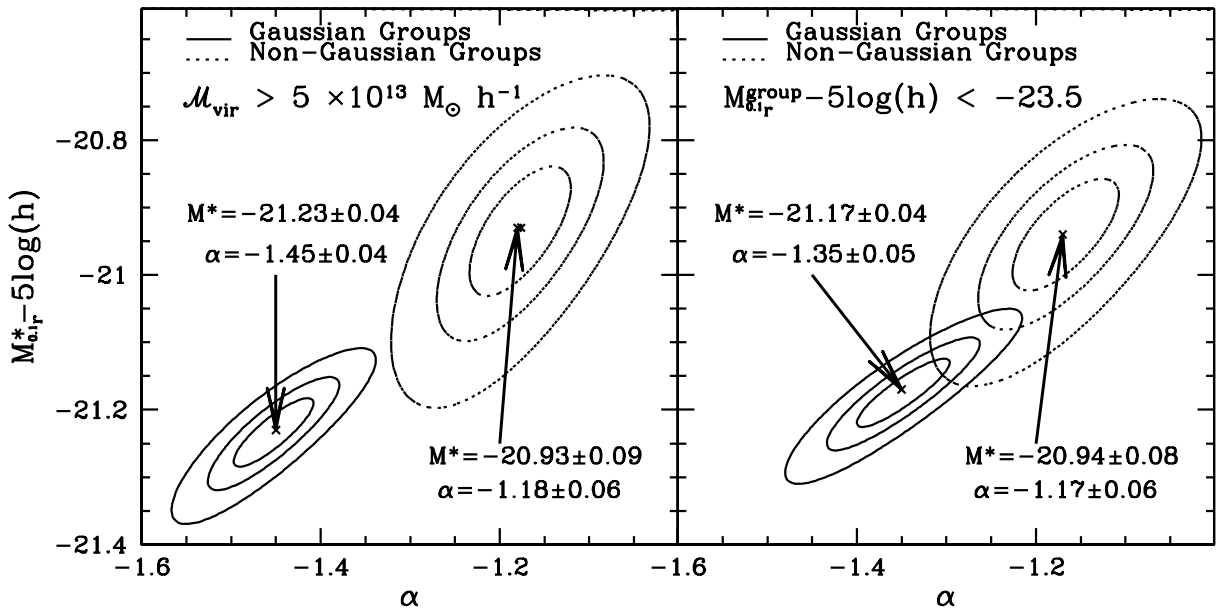}
\caption{
STY best fitting Schechter parameters of the $^{0.1}r$ band LFs of G ({\em solid line})
and NG ({\em dotted line}) groups in the single high mass ({\em left panel}) and
the single high luminosity ({\em right panel}) bins.
We also show their 1, 2 and 3$\sigma$ confidence ellipses.
}
\label{fig5}
\end{figure*}

In agreement with previous results (e.g. \citealt{zmm06,robot10}; ZM11)
there is a clear brightening in the characteristic magnitude 
and  a decreasing faint end slope as a function of mass for both groups subsamples ({\em left
panels}). As expected, the same behaviours are seen as a function of group total absolute magnitude
({\em right panels}). 
Comparing the LF parameters for G and NG groups, we observe that, within
errors, there are no significant differences in the Schechter parameters, with the exception of
the high mass/luminosity tails, where there is an indication that galaxies in G groups 
may have brighter $M^{\ast}$ and steeper $\alpha$ values.  

In order to explore in detail these possible differences 
at the high mass/luminosity tails and to assess their reliability, 
we recompute the LF for massive/luminous groups by using single bins
in mass and luminosity. 
In the single high mass bin we decide to include all groups more massive than  
$5\times 10^{13}M_{\odot}h^{-1}$. With this choice, we include almost completely 
the two highest mass bins of G groups in Fig. \ref{fig3} while keeping a number of NG groups
large enough to have a reliable statistics. 
For the single high luminosity bin, we estimate the corresponding value using the previous high mass 
cut-off and the least square linear relation between group virial mass ($\log({\mathcal M}_{vir})$) 
and group total absolute magnitude ($M_{^{0.1}r}^{\ast}-5\log(h)$) for the ZM11 group sample shown in 
Fig. \ref{fig2}. 
We obtain a group total absolute magnitude cut-off of $-23.5$ in the $^{0.1}r-$band.
As in the virial mass case, this high luminosity subsample includes the two highest luminosity bins 
of G groups in Fig. \ref{fig3}.\\ 
Now, since we are interesting to demonstrate that the dynamical
state of the systems is an indicator of the different luminosity behaviours in the high tails, 
we matched G and NG group distributions of virial mass and group total absolute magnitude in each subsample
(Kolmogorov-Smirnov coefficient $> \ 99.9\%$ in both distributions).
After performing this procedure we obtain 1225 G and 123 NG groups for the high mass subsample and
1373 G and 197 NG groups for the high luminosity subsample.
In the {\em upper panels} of Fig. \ref{fig4} we show the distributions of mass ({\em left panel}) and
group total absolute magnitude ({\em right panel}) for G ({\em solid line}) and NG ({\em dashed line}) groups in 
these high mass/luminosity subsamples. 
We also show in this Figure the number of members ({\em middle panels}) and redshift distributions 
({\em bottom panels}) of G and NG groups for both subsamples.
It can be seen that the NG groups tend have a tail of high membership groups in comparison with the
G group subsample. This tail can be directly associated with the low redshift tail observed for
NG groups.  

In the {\em left panel} of Fig. \ref{fig5} we show the STY
best fitting Schechter parameters for the high mass subsamples of groups along with their
1, 2 and 3 $\sigma$ confidence levels. Clearly, the LF of galaxies in massive G
and NG groups differ at 3$\sigma$ level.
The {\em right panel} of Fig. \ref{fig5} shows the LF parameters corresponding to
G and NG high luminosity groups. Again, we observe a clear difference in the LF of galaxies
in G and NG groups, in this case at a $2\sigma$ significance level.

\section{Discussion}
ZM11 showed that the LF depends not only on local environment (group mass, group-centric distance) 
but also on the large scale environment surrounding the groups. Using the same sample of groups, 
in this work we present evidence that the dynamical state of the system is another important ingredient 
in the evolution of the luminosity of galaxies.
Our results indicate that, for high mass/luminosity groups, the LF of galaxies in G groups have brighter
$M^{\ast}$ and steeper $\alpha$, than the LF of galaxies in NG groups. Therefore, the different 
{\em internal} dynamical state of a system is a clear indicator of a different history in the galaxy 
luminosity evolution.

Systems of galaxies have Gaussian velocity distributions only if they are in dynamical equilibrium. 
Galaxies in these systems have had enough time to suffer the long term action of several physical processes 
during their evolution. Galaxy mergers play a central role
in galaxy evolution in systems. Some other processes such as strangulation
\citep{larson80}, ram pressure \citep{gunn72} and galaxy harassment \citep{farouki81}, are more 
efficient in high mass systems, where the effect observed for different dynamical state of the 
systems it has been shown to be more important. The action of all these processes over the group 
lifetime can produce both, bright and faint galaxies thus providing a plausible explanation for 
our results.

On the other hand, since the stacked velocity distributions for NG groups is well described
by two Gaussian functions, it is likely that the non Gaussianity is caused by the presence of a 
multimodal galaxy population. 
This behaviour opens the possibility for the non Gaussian velocity distributions
to be the consequence of an undergoing merging process (e.g. \citealt{menci96}) or even multiple
merging events (e.g. \citealt{girardi05}).
The different merging populations inhabiting these systems could still be experiencing the
influence of their own parent halo, and hence preserving the galaxy properties 
corresponding to the individual (smaller) halos that are infalling to form the (larger) 
non-Gaussian group. Therefore, these smaller entities should supply the non-Gaussian system 
with less bright galaxy luminosities that correspond to less massive/luminous systems.
This scenario supports the observed fainter characteristic absolute magnitude and the shallower 
faint end slope for non-Gaussian systems.
Galaxies inhabiting these non relaxed systems are unlikely to feel the influence of 
the environmental physical mechanisms described in the previous paragraph, 
thus preventing the formation of very bright galaxies as well as a large number of faint ones.
In agreement with this scenario, \citet{ribeiro10} using the A-D test over a sample of groups from 
the 2PIGG catalogue \citep{eke04} demonstrated that galaxies in Gaussian groups are significantly 
more evolved than galaxies in non Gaussian systems. 
Also, using a subsample of the 2PIGG groups, 
\citet{ribeiro11} have shown that non Gaussian systems are composed of multiple velocity modes, in 
concordance with the scenario of secondary infall of clumps at a stage before virialisation.
Both previous studies were performed analysing the surroundings of groups out to 4 times the corresponding 
radius for an overdensity of 200 ($4R_{200}$). 
In our work we show that a similar behaviour can be observed from the analysis of the internal dynamics 
of groups (only galaxy members, mostly inside the virial radius) and these different dynamical 
environments can be evidenced in the galaxy luminosities of high mass/luminous systems.
Our results suggest another way to test models of galaxy evolution, since the connection between
galaxy luminosities (i.e. astrophysics) and the dynamics of the systems should be present. 

\section*{Acknowledgements}
{\small We thanks to the referee for helpful comments and suggestions that improved the paper.
This work has been supported by Consejo 
Nacional de Investigaciones Cient\'\i ficas y T\'ecnicas de la Rep\'ublica 
Argentina (CONICET, PIP2011/2013 11220100100336),
Secretar\'\i a de Ciencia y Tecnolog\'\i a de la Universidad de C\'ordoba, Argentina.
(SeCyT) and Funda\c c\~ao de Amparo \`a Pesquisa do Estado do S\~ao Paulo
(FAPESP), Brazil.

Funding for the SDSS and SDSS-II has been provided by the Alfred P. Sloan Foundation, the Participating Institutions, the National Science Foundation, the U.S. Department of Energy, the National Aeronautics and Space Administration, the Japanese Monbukagakusho, the Max Planck Society, and the Higher Education Funding Council for England. The SDSS Web Site is http://www.sdss.org/.
The SDSS is managed by the Astrophysical Research Consortium for the Participating Institutions. The Participating Institutions are the American Museum of Natural History, Astrophysical Institute Potsdam, University of Basel, University of Cambridge, Case Western Reserve University, University of Chicago, Drexel University, Fermilab, the Institute for Advanced Study, the Japan Participation Group, Johns Hopkins University, the Joint Institute for Nuclear Astrophysics, the Kavli Institute for Particle Astrophysics and Cosmology, the Korean Scientist Group, the Chinese Academy of Sciences (LAMOST), Los Alamos National Laboratory, the Max-Planck-Institute for Astronomy (MPIA), the Max-Planck-Institute for Astrophysics (MPA), New Mexico State University, Ohio State University, University of Pittsburgh, University of Portsmouth, Princeton University, the United States Naval Observatory, and the University of Washington.}


\label{lastpage}

\end{document}